\documentclass{article}
\usepackage{latexsym}
\usepackage{jgaa-article}
\usepackage{graphicx}
\usepackage{url}

\newtheorem{lemma}{Lemma}
\newtheorem{theorem}{Theorem}
\newtheorem{defn}{Definition}
\newtheorem{corollary}{Corollary}
\newenvironment{proof}{\par\addvspace\topsep\noindent
{\bf Proof:} \ignorespaces }{\qed}
\newcommand{\qed}{\hspace*{\fill}$\Box$\ifmmode\else\par\addvspace\topsep\fi}

\newdimen\pIR
\newcommand\R{{\rm I\kern\pIR R}}

\begin{document}

\Issue{3}{3}{1--27}{1999}
\HeadingAuthor{D. Eppstein}
\HeadingTitle{Planar Subgraph Isomorphism}

\Title{Subgraph Isomorphism in Planar Graphs and Related Problems}

\Author{David Eppstein}
\Institution{Department of Information and Computer Science\\
University of California, Irvine\\
\Web{http://www.ics.uci.edu/$\sim$eppstein/}\\
\Email{eppstein@ics.uci.edu}}

\begin{abstract}
We solve the subgraph isomorphism problem in planar graphs in linear
time, for any pattern of constant size.  Our results are based on a
technique of partitioning the planar graph into pieces of small
tree-width, and applying dynamic programming within each piece.  The
same methods can be used to solve other planar graph problems including
connectivity, diameter, girth, induced subgraph isomorphism, and
shortest paths.
\end{abstract} 

\History{Communicated by Roberto Tamassia: submitted December 1995,
revised November~1999.}

\Ack{Work supported in part by NSF grant CCR-9258355
and by matching funds from Xerox Corp.}

\Body

\section{Introduction}

Subgraph isomorphism is an important and very general form of exact
pattern matching.  Subgraph isomorphism is a common
generalization of many important graph problems including finding
Hamiltonian paths, cliques, matchings, girth, and shortest paths.
Variations of subgraph isomorphism have also been used to model such
varied practical problems as molecular structure
comparison~\cite{ArtBatGri-JCICS-92}, integrated circuit
testing~\cite{BroTho-IEEI-88}, microprogrammed controller
optimization~\cite{Guh-ICCD-87},
prior-art avoidance in genetic evolution of
circuits~\cite{KozBenSti-EGP-99},
analysis of Chinese
ideographs~\cite{DonWuDin-ICPR-88}, robot motion
planning~\cite{LanWon-ISIC-91}, semantic network
retrieval~\cite{Lev-CMA-92}, and polyhedral object
recognition~\cite{StaWah-CAI-92}.

In the subgraph isomorphism problem, given a ``text'' $G$ and a
``pattern'' $H$, one must either detect an occurrence of $H$ as a
subgraph of $G$, or list all occurrences.  For certain choices of $G$
and $H$ there can be exponentially many occurrences, so listing all
occurrences can not be solved in subexponential time.  Further, the
decision problem is NP-complete.  However for any fixed pattern $H$ with
$\ell$ vertices, both the enumeration and decision problems can easily
be solved in polynomial $O(n^\ell)$ time, and for some patterns an even
better bound might be possible.  Thus one is led to the problem of
determining the algorithmic complexity of subgraph isomorphism for a
fixed pattern.

Here we consider the special case in which $G$ (and therefore $H$) are
planar graphs, a restriction naturally occurring in many applications.  We
show that for any fixed pattern, planar subgraph isomorphism can be
solved in linear time.  Our results extend to some other problems
including vertex connectivity, induced subgraph isomorphism and shortest
paths.

Our algorithm uses a graph decomposition method similar to one
used by Baker~\cite{Bak-JACM-94} to approximate various NP-complete
problems on planar graphs.  Her method involves removing vertices from
the graph leaving a disjoint collection of subgraphs of small tree-width;
in contrast we find a collection of non-disjoint subgraphs of small
tree-width covering the neighborhood of every vertex.

We assume throughout that all planar graphs are {\em simple},
so that the number of edges is at most $O(n)$; this simplifies our time
bounds as we need not include the dependence on this number.
The only problems for which this assumption makes a difference
are induced subgraph isomorphism, $h$-clustering, and edge connectivity;
for those, one can assume without loss of generality
that the graph has bounded edge multiplicity, so again $m=O(n)$.

\section{New Results}

We prove the following results.
The time dependence on $H$ is omitted from these bounds.
In general it is exponential
(necessarily so, unless P=NP, since planar subgraph isomorphism is
NP-complete) but see Theorem~\ref{Hamiltonian}
for situations in which it can be improved.

\begin{itemize}
\item
We can test whether any fixed pattern~$H$ is a subgraph of a planar graph $G$,
or count the number of occurrences of $H$ as a subgraph of $G$, in time~$O(n)$.

\item If {\em connected} pattern~$H$
has $k$ occurrences as a subgraph of a planar graph $G$, we can list
all occurrences in time $O(n + k)$.  If $H$
is 3-connected, then $k=O(n)$~\cite{Epp-JGT-93}, and we can list all
occurrences in time~$O(n)$.

\item
We can count the number of {\em induced} subgraphs of a planar graph~$G$
isomorphic to any fixed connected pattern $H$ in time $O(n)$, and if
there are $k$ occurrences we can list them in time $O(n+k)$.

\item For any planar graph $G$ for which we know a constant bound on the
diameter, we can compute the exact diameter in time $O(n)$.

\item For any constant $h$ we can solve the $h$-clus\-ter\-ing and
connected
$h$-clus\-ter\-ing problems~\cite{KeiBre-CMCC-91} in planar graphs in time
$O(n)$.

\item For any planar graph $G$ for which we know a constant bound on the
girth, we can compute the exact girth in time $O(n)$.
The same bound holds if instead of girth we ask for the shortest
separating cycle or for the shortest nonfacial cycle in a given plane
embedding of the graph.

\item For any planar graph $G$, we can compute the vertex
connectivity and edge connectivity of $G$ in time $O(n)$.
(For planar multigraphs, we can test $k$-edge-connectivity for
any fixed $k$ in time $O(n)$.)

\item For any planar graph $G$ and any constant $\ell$,
we construct in time $O(n)$ a linear-space routing data structure which
can test for any pair of vertices whether their distance is at most
$\ell$, and if so find a shortest path between them, in time $O(\log n)$.
\end{itemize}

\section{Related Work}

For general subgraph isomorphism, nothing better than the
naive exponential $O(n^{|H|})$ bound is known.  Plehn and
Voigt~\cite{PleVoi-WG-90} give an algorithm for subgraph isomorphism
which in planar graphs takes time $|H|^{O(|H|)}n^{O(\sqrt{|H|})}$
(since improved by Alon et al.~\cite{AloYusZwi-JACM-95} to
$2^{O(|H|)}n^{O(\sqrt{|H|})}$), but this is still much
larger than the linear bound we achieve.

Several papers have studied planar subgraph isomorphism with restricted
patterns.  It has long been known that if the pattern~$H$ is either
$K_3$ or $K_4$, then there can be at most $O(n)$ instances of $H$ as a
subgraph of a planar graph $G$, and that these instances can be listed
in linear time~\cite{BarEve-STOC-82,ItaRod-SJC-78,PapYan-IPL-81}, a fact
which has been used in algorithms to test connectivity~\cite{Lau-IPL-90},
to approximate maximum independent sets~\cite{BarEve-STOC-82}, and to test
inscribability~\cite{DilSmi-SCG-92}. Linear time and instance bounds for
$K_3$ and
$K_4$ can be shown to follow solely from the sparsity properties of planar
graphs~\cite{ChiNis-SJC-85,ChrEpp-TCS-91}, and similar methods also
generalize to problems of finding $K_{2,2}$ and other complete bipartite
subgraphs~\cite{ChiNis-SJC-85,Epp-IPL-94}.  Richards~\cite{Ric-Algs-86}
gives
$O(n\log n)$ algorithms for finding $C_5$ and $C_6$ subgraphs in planar
graphs, and leaves open the question for larger cycle lengths; Alon et
al.~\cite{AloYusZwi-JACM-95} gave $O(n\log n)$ deterministic and $O(n)$
randomized algorithms for larger cycles.  In~\cite{Epp-JGT-93}, we showed
how to list all cycles of a given fixed length in {\em outerplanar}
graphs, in linear time (see
also~\cite{Lin-TCS-89,LinPro-TCS-89,LinSys-ICALP-88,Sys-TCS-82} for
similar variants of outerplanar subgraph isomorphism).  We used our
outerplanar cycle result to find any {\em wheel} of a given fixed size in
planar graphs, in linear time.  Itai and Rodeh~\cite{ItaRod-SJC-78}
discuss the problem of finding the girth of a general graph, or
equivalently that of finding short cycles.  The problem of finding cycles
in planar graphs was discussed above.  Fellows and
Langston~\cite{FelLan-STOC-89} discuss the related problem of finding a
path or cycle longer than some given length in a general graph, which
they solve in linear time for a given fixed length bound.  The planar
dual to the shortest separating cycle problem has been related by Bayer
and Eisenbud~\cite{BayEis-AM-91} to the Clifford index of certain
algebraic curves.  Our results here generalize and unify this collection
of previously isolated results, and also give improved dependence on the
pattern size in certain cases.

Recently we were able to characterize the graphs that can occur at most
$O(n)$ times as a subgraph isomorph in an $n$-vertex planar graph:
they are exactly the
3-connected planar graphs~\cite{Epp-JGT-93}.  However our proof does not
lead to an efficient algorithm for 3-connected planar subgraph
isomorphism.  In this paper we use different techniques which do not
depend on high-order connectivity.

Laumond~\cite{Lau-IPL-90} gave a linear time algorithm
for finding the vertex connectivity of maximal planar graphs.
Eppstein et al.~\cite{EppGalIta-SJC-99} give an $O(n)$ time
algorithm for testing $k$-edge-connectivity for $k\le 4$ and
$k$-vertex-connectivity for $k\le 3$.
For general graphs, testing $k$-edge-connectivity for fixed $k$
takes time $O(m + n\log n)$~\cite{Gab-JCSS-95}.
4-vertex-connectivity in general graphs can be tested in time
$O(n\alpha(n)+m)$~\cite{KanTamDiB-FOCS-91}.  However planar graphs can
be as much as 5-vertex-connected, and nothing even close to linear
was known for testing planar 5-connectivity.

Our shortest path data structure combines our methods of bounded
tree-width decomposition with a separator-based divide and conquer
technique due to Frederickson~\cite{Fre-SJC-87}.  Obviously all pairs
shortest paths can be computed in time $O(nm)$ after which the queries
we describe can be answered in time $O(1)$, but some faster algorithms
are known for approximate planar shortest
paths~\cite{FreJan-SJC-89,FreJan-SJC-90,KleSub-Algo-98}. Our data
structure answers shortest path queries exactly, in less preprocessing
time than the other known results, but can only find paths of constant
length.

A final note of caution is in order.  One should not be confused by the
superficial similarity between the subgraph isomorphism problems posed here
and the graph minor problems studied extensively by Robertson,
Seymour, and others \cite{GST-91}. One can recognize path subgraphs by
minor testing, but such tricks do not work for most other subgraph
isomorphism problems. The absence of a fixed {\em minor} imposes severe
structural constraints on a graph, whereas this is much less the case
when a fixed {\em subgraph} is not present. Although minor testing can be
done in time polynomial in the text graph size, the constant factors are
typically much higher than those for our subgraph isomorphism algorithm.

\section{Bounded Tree-Width Subgraph Isomorphism}

\begin{figure}[t]
$$\includegraphics[width=4.5in]{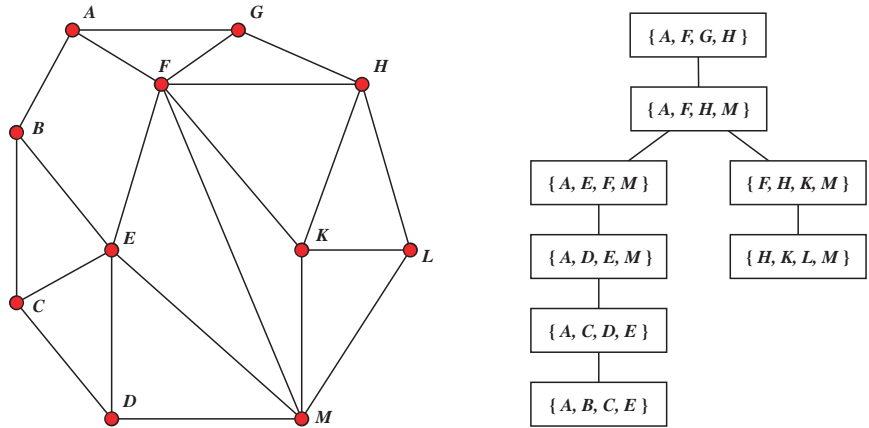}$$
\caption{Tree decomposition of a planar graph.}
\label{fig:treedec}
\end{figure}

As a subroutine, we need to perform subgraph isomorphism testing in
graphs of bounded tree-width.  This can be done by a standard dynamic
programming technique~\cite{BerLawWon-Algs-87,TakNisSai-JACM-82}.  The
exact statement of the problem we solve is complicated by the requirement
that we count or list each subgraph isomorph exactly once.  For
simplicity, we state the bounds for this problem with one parameter
measuring both the tree-width of the text and the size of the pattern.

\begin{defn}
A {\em tree decomposition} of a graph $G$ consists of a tree $T$,
in which each node $N\in T$ has a {\em label} $L(N)\subset V(G)$,
such that the set of tree nodes whose labels contain any particular
vertex of $G$ forms a contiguous subtree of $T$, and such that any edge of
$G$ connects two vertices belonging to the same label $L(N)$ for at
least one node $N$ of $T$.  The {\em width} of the tree decomposition is 
one less than the size of the largest label set in $T$. The {\em
tree-width} of $G$ is the minimum width of any tree decomposition of $G$.
\end{defn}

We can assume without loss of generality (by splitting high-degree nodes
into multiple nodes with the same label) that each node in $T$ has at most
three neighbors and that there are $O(n)$ nodes in the tree. We will
assign our tree decomposition an arbitrary orientation, by rooting it at
one of its leaves, so that $T$ becomes a binary tree.
Figure~\ref{fig:treedec} shows a planar graph, with a tree decomposition
of width three.  In fact, the graph shown has no tree decomposition of
width two, so its tree-width is three.

Define the subtree rooted at a node $N$ to consist of $N$ and all its
descendants. Each such subtree is associated with an induced subgraph of
$G$, having vertices contained in labels of nodes in the subtree.

\begin{lemma}
The subtree rooted at $N$ provides a tree decomposition of the
associated induced subgraph of $G$.
\end{lemma}

\begin{proof}
The only property of a tree decomposition that does not follow
immediately is the requirement that each edge connect two vertices
contained in the label of some node.  Since this is true of $G$ and $T$,
any induced subgraph edge $(u,v)$ must have $\{u,v\}\subset L(N')$
for some $N'$, but $N'$ may not be a descendant of $N$.  However, if not, 
$u$ belongs to both $L(N')$ and (by assumption) $L(N'')$ where $N''$ is a
descendant of
$N$.  Therefore, by contiguity, $u\in L(N)$, and similarly $v\in L(N)$,
so in this case $(u,v)$ still both belong to
the label of at least one node in the subtree.
\end{proof}

\begin{figure}[t]
$$\includegraphics[width=4.5in]{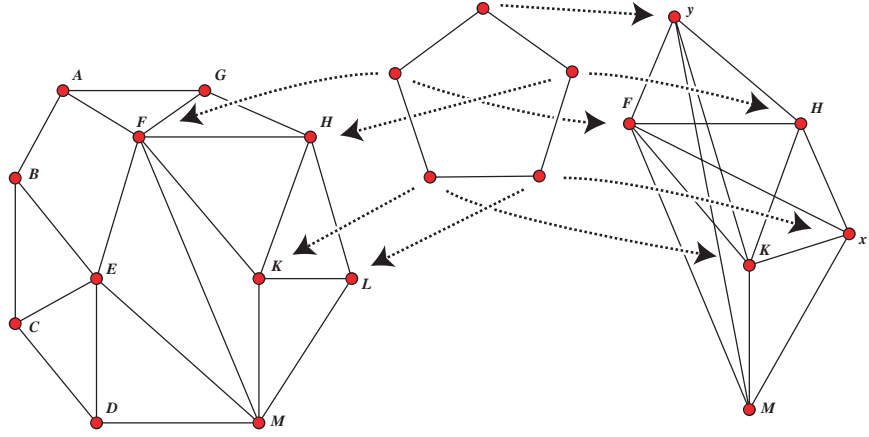}$$
\caption{Partial isomorph of a pentagon in the induced subgraph
associated with node $\{F,H,K,M\}$, and corresponding
partial isomorph boundary mapping the pentagon to $G'_N$.}
\label{fig:partiso}
\end{figure}

\begin{lemma}\label{dyn-prog-count}
Assume we are given graph $G$ with $n$ vertices along with a tree
decomposition $T$ of $G$ with width $w$.  Let $S$ be a subset of the
vertices of
$G$, and let $H$ be a fixed
graph with at most $w$ vertices.  Then in time $2^{O(w\log w)}n$ we can
count all isomorphs of $H$ in $G$ that include some vertex in $S$.  We
can list all such isomorphs in time
$2^{O(w\log w)}n + O(kw)$, where $k$ denotes the
number of isomorphs and the term $kw$ represents the total output size.
\end{lemma}

\begin{proof}
We perform dynamic programming in tree $T$.
Let a {\em partial isomorph} at a node $N$ of the tree
be an isomorphism between an induced subgraph $H'$ of the pattern $H$
and the induced subgraph of $G$ associated with the subtree
rooted at $N$.

We let $G'_N$ be formed by adding two additional vertices $x$, $y$
to the subgraph of $G$ induced by vertex set $L(N)$.  We connect each of
the two additional vertices to all vertices in $L(n)$, and each
of the two additional vertices also is given a self-loop. Then from any
partial isomorph at $N$ we can derive a graph homomorphism from all of
$H$ to $G'$,  which is one-to-one on vertices in $L(N)$,
maps the rest of $H'$ to $x$, and maps $H-H'$ to
$y$. Let a {\em partial isomorph boundary} be such a map;
Figure~\ref{fig:partiso} illustrates a partial isomorph and the
corresponding boundary. Since a partial isomorph boundary consists of a
map from a set of at most $w$ objects to a set of at most $w+3$ objects,
there are at most
$w^{w+3}=2^{O(w\log w)}$ possible partial isomorph boundaries for a
given node.

Suppose that node $N$ has children $N_1$ and $N_2$
We say that two partial isomorph boundaries $B:H\mapsto G'_N$
and  $B_1:H\mapsto G'_{N_1}$
are {\em consistent} if 
the following conditions all hold:
\begin{itemize}
\item For each vertex $v\in H$, if $B(v)\in L(N_1)$ or
$B_1(v)\in L(N)$, then $B(v)=B_1(v)$.
\item For each vertex $v\in H$, if $B(v)\ne x$ then
$B_1(v)\in L(N)\cup\{y\}$.
\item At least one vertex $v\in H$ has $B_1(v)\not\in L(N)\cup\{y\}$.
\end{itemize}
We say that two partial isomorph boundaries $B_1:H\mapsto G'_{N_1}$
and $B_2:H\mapsto G'_{N_2}$ form a {\em compatible triple} with $B$
if the following conditions both hold:
\begin{itemize}
\item $B_1$ and $B_2$ are both consistent with $B$.
\item For each $v$ with $B(v)=x$, exactly one of
$B_1(v)$ and $B_2(v)$ is equal to $y$.
\end{itemize}

For each partial isomorph boundary $B:H\mapsto G'_N$,
let $X_1(B)$ be the
number of partial isomorphs which give rise to that boundary,
and include a vertex of $S$.
Let $X_2(B)$ be the number of partial isomorphs which give rise to that
boundary, and do not include a vertex of $S$.
These values can be computed in a bottom-up fashion as follows:

\begin{itemize}
\item If there is no $v$ for which $B(v)=x$,
then all partial isomorphs having boundary $B$ 
involve only vertices in $L(N)$, and can be enumerated by brute force
in time $2^{O(w\log w)}$.

\item Otherwise, we initialize $X_1(B)$ and $X_2(B)$ to zero.
Then, for each partial boundary $B_1$ that is consistent with
$B$, and such that there is no $v$ with $B(v)=x$ and $B_1(v)=y$,
we increment $X_1(B)$ by $X_1(B_1)$ and increment
$X_2(B)$ by $X_2(B_1)$.
Finally, for each compatible triple $B, B_1, B_2$
we increment $X_1(B)$ by $X_1(B_1)\cdot X_1(B_2)+X_1(B_1)\cdot
X_2(B_2)+X_2(B_1)\cdot X_1(B_2)$
and increment $X_2(B)$ by $X_2(B_1)\cdot X_2(B_2)$.
\end{itemize}

The total time for testing all triples for compatibility and performing
the above computation is $O(w^{3(w+3)+1}=2^{O(w\log w)}$.

At the root node of the tree, we compute the number of isomorphs
involving $S$ simply by summing the values
$X_2(B)$ over all partial isomorph boundaries for which $B(v)\ne y$ for
all $v$.
\end{proof}

\begin{lemma}\label{dyn-prog-list}
Assume we are given graph $G$ with $n$ vertices along with a tree
decomposition $T$ of $G$ with width $w$.  Let $S$ be a subset of the
vertices of
$G$, and let $H$ be a fixed
graph with at most $w$ vertices.  Then we can list all isomorphs of $H$
in $G$ that include some vertex in $S$
in time
$2^{O(w\log w)} n + O(kw)$, where $k$ denotes the
number of isomorphs and the term $kw$ represents the total output size.
\end{lemma}

\begin{proof}
We first follow the above dynamic programming procedure,
to compute the values $X_1$ and $X_2$ for each partial isomorph boundary.
We then compute top-down in the tree the set of
pairs $(B,X)$ where $B$ is a partial isomorph boundary and $X$ is either
$X_1$ or $X_2$, such that the value $X(B)$ contributes to the final
count of subgraph isomorphs.  These pairs can be identified as the ones
such that $X(B)$ was included in the computation of some pair higher in
the tree that has been previously identified as contributing to the
total, and that caused a nonzero increment in this computation.
Finally, we compute bottom-up again, listing for each contributing pair
$(B,X)$ the partial subgraph isomorphs counted in the value $X(B)$.
This step can be performed by mimicking the initial computation of
$X(B)$ described in the previous lemma, restricted to the boundaries
known to contribute to the overall total, replacing each increment by
a concatenation of lists, and replacing each multiplication with the
construction of partial isomorphs from a Cartesian product of two
previously-computed lists.

The number of steps for this computation is proportional to the number
of steps in the previous algorithm, together with the added time for
each combination of a pair of partial isomorphs.  Each such combination
can be charged to a subgraph isomorph included in the output, and each
output isomorph is formed by a binary tree of combinations that takes
$O(w)$ time to perform, so the total added time is $O(kw)$.
\end{proof}

The same dynamic programming techniques also lead to similar results for
counting or listing induced subgraphs isomorphic to $H$.  To do this, we
need only modify the algorithms above to restrict attention to partial
isomorph boundaries $B:H\mapsto G'_N$ in which all edges between
vertices of $L(N)$ are covered by the image of some edge in $H$.

\section{Neighborhood Covers}

We have seen above that we can perform subgraph isomorphism quickly in graphs
of bounded tree-width.  The connection with planar graphs is the following:

\begin{figure}[t]
$$\includegraphics[width=4.5in]{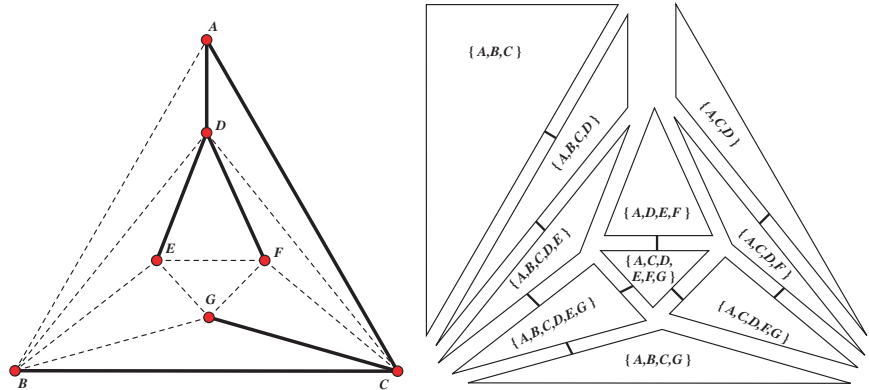}$$
\caption{Triangulated planar graph, with depth two tree $T$ rooted at $A$
(shown by heavy solid lines), and tree decomposition with nodes
corresponding to faces of the graph and edges complementary to $T$.}
\label{fig:bfswidth}
\end{figure}

\begin{lemma}[Baker~\cite{Bak-JACM-94}]\label{diam-width}
Let planar graph $G$ have a rooted spanning tree $T$ in
which the longest path has length $\ell$.  Then
a tree decomposition of $G$ with width at most $3\ell$ can be found in
time
$O(\ell n)$.
\end{lemma}

\begin{proof}
Without loss of generality (by adding edges if necessary) we can assume
$G$ is embedded in the plane with all faces
triangles (including the outer face). Form a tree with one node per
triangle, and an edge connecting any two nodes whenever the
corresponding triangles share an edge that is not in $T$
(Figure~\ref{fig:bfswidth}).
Label each node with the set of vertices on the paths connecting each
corner of the triangle to the root of the tree.
Then each edge's endpoints are part of some label set (namely, the sets
of the two triangles containing the edge), and the labels containing any
vertex form a contiguous subtree (namely, the path of triangles
connecting the two triangles containing the edge from the vertex to its
parent, and any other triangles enclosed in the embedding by this path).
Therefore, this gives us a tree decomposition of
$G$. The number of nodes in any label set is at most $3\ell+1$,
so the width of the decomposition is at most $3\ell$.
\end{proof}

In particular, any planar graph with diameter $D$ has tree-width $O(D)$.

If an isomorph of a connected pattern $H$ uses vertex $v$ in $G$, it is
contained in the portion of $G$ within distance $|H|$ of $v$.  By
Lemma~\ref{diam-width} this {\em $|H|$-neighborhood} of $v$ has tree-width
at most $3|H|$.  Therefore we can {\em cover} $G$ by the collection of all
such neighborhoods, and use Lemma~\ref{dyn-prog-list} to find the copies
of
$H$ within each neighborhood.  However such a cover is not {\em
efficient}: the total size of all subgraphs is $O(n^2)$, so this would
give us a subgraph isomorphism algorithm with quadratic runtime.  We
speed this up to linear by using more efficient covers.

Awerbuch et al.~\cite{AweBerCow-SWAT-92,AwePel-FOCS-90} have introduced
the very similar concept of a {\em neighborhood cover}, which is a
covering of a graph by a collection of subgraphs, with the properties
that the neighborhood of every vertex is contained in some subgraph, and
that every subgraph has small diameter.  They showed that for any
(possibly nonplanar) graph, and any given value $w$, there is a
$w$-neighborhood cover in which the diameter of each subgraph is
$O(w\log n)$, and in which the total size of all subgraphs is $O(m\log
n)$; such a cover can be computed in time $O(m\log n + n\log^2
n)$~\cite{AweBerCow-SJC-98}. Because of Lemma~\ref{diam-width}, such a
neighborhood cover is also almost exactly what we want to speed up our
subgraph isomorphism algorithm.  However there are two problems.  First,
the size and construction time of neighborhood covers are higher than we
want (albeit only by polylogarithmic factors). Second, and more
importantly, the diameter of each subgraph is logarithmic, so we are
unable to use dynamic programming directly in the subgraphs of the
cover.  We would instead be forced to use some additional techniques such
as separator-based divide and conquer, introducing more unwanted
logarithmic factors.

Instead, we use a technique similar to that of Baker~\cite{Bak-JACM-94}
to form a cover that has the properties we want directly:
any connected $w$-vertex subgraph of $G$ is included in
some member of the cover,
and each vertex of $G$ is included in few members of the cover
(so the total size of the cover is $O(n)$).
Unlike the techniques cited above, the diameter of the subgraphs will not
be bounded, however we will still be able to use Lemma~\ref{diam-width}
on an auxiliary graph to show that each covering subgraph has
tree-width~$O(w)$.  Because of the exponential dependence of our overall
algorithms on the tree-width of the covering subgraphs, we concentrate
our efforts on reducing this width as much as possible, at the expense
of increasing the total size of the cover by an $O(w)$ factor over the
minimum possible.

\begin{lemma}\label{nbhd-cover}
Let $G$ be a planar graph, and $w$ be a given integer parameter.  Then
we can find a collection of subgraphs $G_i$ and a partition of the
vertices of $G$ into subsets $S_i$ with the following properties:
\begin{itemize}
\item Every vertex of $G$ is included in at most $w$ subgraphs $G_i$.

\item We can find a tree decomposition of each
subgraph $G_i$ with tree-width at most $3w-1$.

\item If $H$ is a connected $w$-vertex subgraph of $G$,
and $i$ is the smallest value for which $H\cap S_i$ is nonempty,
then $H$ is a subgraph of $G_i$
but is not a subgraph of any $G_j$ with $j>i$.

\item The total time for performing the partition and computing the tree
decompositions is $O(w^2 n)$.
\end{itemize}
\end{lemma}

\begin{figure}[t]
$$\includegraphics[width=4in]{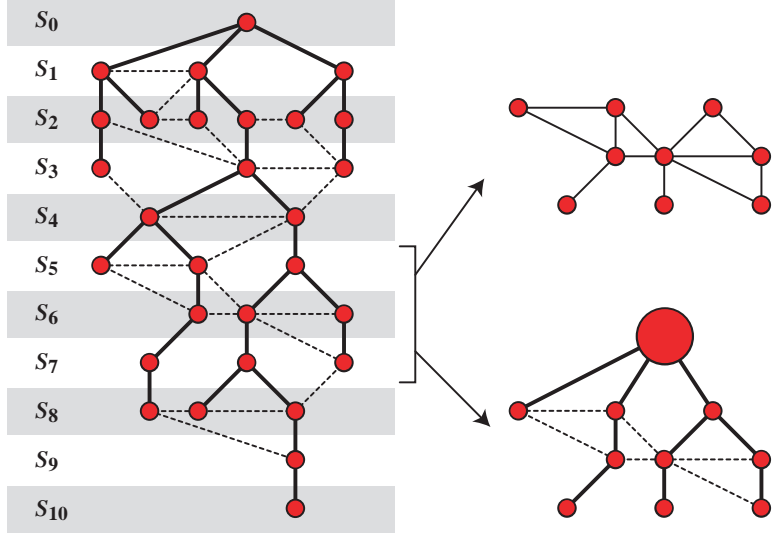}$$
\caption{Planar graph with breadth first spanning tree (heavy edges),
partition into layers $S_i$, subgraph $G_i$ (for $w=3$, $i=5$),
and minor $G'_i$ (with large supervertex and contracted breadth first
spanning tree).}
\label{fig:subcover}
\end{figure}

\begin{proof}
We choose an arbitrary starting vertex $v_0$, and let $S_i$
consist of the vertices at distance $i$ from $v_0$.
We then let $G_i$ be the graph induced
by the vertex set $\bigcup_{j=i}^{i+w-1} S_j$,
as shown in Figure~\ref{fig:subcover}.
Clearly, the sets $S_i$ form a partition of the vertices of $G$,
and each vertex is in at most $w$ subgraphs $G_i$.

Then for $i=0$, $G_i$ consists of
the vertices at distance at most $w-1$ from $v_0$, so by
applying Lemma~\ref{diam-width} to its breadth first spanning tree
we can find a tree decomposition with width at most $3(w-1)$.
To show that each $G_i$ with $i>0$ has low tree-width, form an auxiliary
graph
$G'_i$ from $G$ by collapsing into a single supervertex all the vertices
at distance less than $i$ from $v_0$, and deleting all the vertices with
distance at least $i+w$.  $G'_i$ is a minor of the planar graph $G$
and is therefore also planar.
Then similarly collapsing a breadth first spanning
tree of $G$ gives a spanning tree of $G'_i$ with depth at most $w$,
so $G'_i$ has a tree decomposition with width at most $3w$,
in which each node of the decomposition includes the collapsed
supervertex in its label.  $G_i$ is formed by deleting this supervertex
from $G'_i$, so we can form a tree decomposition of $G_i$ with width at
most $3w-1$ by removing the supervertex from the decomposition of $G'_i$.

Next, we need to show that any connected subgraph $H$ of $G$
with $|H|\le w$ is contained in $G_i$, where $i$ is the smallest value
such that $H\cap S_i\ne\emptyset$.  But $G_i$ is formed from $G$ simply
by removing the sets $S_j$ where $j<i$ or $j\ge i+w$.
No $S_j$ with $j<i$ can contain a vertex of $H$, or else $i$ would have
been smaller.  And no $S_j$ with $j\ge i+w$ can contain a vertex $v$ of
$H$, or else we could find a path of length at most $i+w-1$ from $v_0$ to
$v$ by concatenating a path in $H$ from some vertex $v_i\in S_i\cap H$
to $v$ (which has length at most $|H|-1\le w-1$) with the breadth first
tree path from $v_0$ to $v_i$ (which has length $i$), contradicting the
placement of $v$ in $S_j$.  Therefore, none of the vertices that were
deleted from $G$ can belong to $H$, so $H$ remains a subgraph of $G_i$.

Finally, the condition that $H$ can not be a subgraph for $G_j$ where
$j>i$ is clearly true, since no such $G_j$ can include any vertex of
$S_i$.

The time bound is dominated by the time to perform the tree
decompositions on the graphs $G'_i$, which by Lemma~\ref{diam-width} is
$\sum O(w G_i) = w O(\sum G_i) = w(wn)$.
\end{proof}

\section{The Subgraph Isomorphism Algorithm}

We first describe the result for the special case of connected patterns.

\begin{theorem}\label{main}
We can count the isomorphs or induced isomorphs of a given connected
pattern $H$, having $w$ vertices, in a planar text graph $G$ with $n$
vertices, in time $2^{O(w\log w)} n$.  If there are $k$ such isomorphs
we can list them all in time $2^{O(w\log w)} n + O(wk)$.
\end{theorem}

\begin{proof}
The algorithm consists of the following steps:
\begin{enumerate}
\item Apply the method of Lemma~\ref{nbhd-cover}
to find a partition of the vertices into sets $S_i$
associated with graphs $G_i$ having low width tree decompositions.
\item For each $i\ge 0$, count or list the subgraph isomorphs of $H$ in
$G_i$ that involve at least one vertex of $S_i$, using the algorithm
of Lemma~\ref{dyn-prog-count} or Lemma~\ref{dyn-prog-list} respectively.
\item Sum all the counts or concatenate the lists, to get a count or
list of the isomorphs in $G$.
\end{enumerate}
By Lemma~\ref{nbhd-cover}, each isomorph of $H$ in $G$
occurs in exactly one way as an isomorph in $H_i$ that involves at least
one vertex of $S_i$, so the algorithm produces the correct total count
or list.
The time for the first step is $O(w^2 n)$, and the time for the last
step is $O(n)$, both dominated by the time for the second step which is
$\sum 2^{O(w\log w)}|G_i| + O(k_i w)$,
where the $2^{O(w\log w)}$ factor arises by plugging the $3w-1$ treewidth
bound of Lemma~\ref{nbhd-cover} into the analysis in Lemmas
\ref{dyn-prog-count} and~\ref{dyn-prog-list}.
This can be simplified to
$2^{O(w\log w)}(\sum |G_i|)  + O(\sum k_i w)
= 2^{O(w\log w)}\cdot O(wn)+O(kw) = 2^{O(w\log w)}n+O(kw)$.
\end{proof}

The method so far requires that the pattern be connected.
We now describe a general method for handling disconnected patterns.
The technique will let us count the number of matching patterns,
after which some sort of separator-based divide and conquer can likely
be used to find an instance of a matching pattern, but we have been
unable to extend this technique to the problem of listing all subgraph
isomorphs of a disconnected pattern.

\begin{theorem}\label{disconnected}
We can count the isomorphs of any (possibly disconnected) pattern $H$
having at most $w$ of vertices, in a planar text graph $G$ with
$n$ vertices,
in time $2^{O(w\log w)} n$.
\end{theorem}

\begin{proof}
Let $\#_G(H)$ denote the number of isomorphs of $H$ in $G$.
Rather than counting the isomorphs of a single pattern, we
count the isomorphs of all planar graphs having at most $w$
vertices.  There are only $2^{O(w)}$ such graphs~\cite{Tur-DAM-84},
so this factor does not change the overall form of our time bound.
We order these graphs by the number of connected components,
so that when we are processing a particular graph $H$ we can assume we
already know the values of $\#_G(H')$ for every $H'$ with fewer
components.

Our algorithm them performs the following steps on each graph $H$:
\begin{enumerate}
\item If $H$ is connected, compute $\#_G(H)$ using the algorithm
of Theorem~\ref{main}.
\item Otherwise, let $H$ be the disjoint union of two subgraphs
$H'$ and $H''$, and let
$\#_G(H) = \#_G(H')\cdot\#_G(H'') - \sum k_i\#_G(H_i)$,
where the sum is over all graphs $H_i$ with fewer components than
$H$, and $k_i$ denotes the number of different ways $H_i$ can be formed
as the union of $H_1$ and $H_2$.
\end{enumerate}

The product $\#_G(H')\cdot\#_G(H'')$
counts the number of ways of mapping $H$ into $G$ such that
both $H'$ and $H''$ are isomorphically mapped but their instances may
overlap.  The term $\sum k_i\#_G(H_i)$ corrects for these overlaps
by subtracting the number of overlapped occurrences of each possible
type.

The coefficients $k_i$ may be computed by brute force enumeration
of all possible ways of marking a vertex of $H_i$ as coming from $H'$,
$H''$, or both, combined with a planar graph isomorphism algorithm,
in time $2^{O(w)}$.  
Therefore, the overall time taken in the second step of the algorithm
is $2^{O(w)}$, independent of $n$, and the
total time is dominated by the first step,
in which we apply Theorem~\ref{main} to $2^{O(w)}$ connected graphs,
taking time $2^{O(w\log w)} n$.
\end{proof}

\section{Further Improvements}

For certain patterns, such as the wheels, our results can be further
improved to reduce the time dependence on $|H|$.
Let ${\rm diam}(H)$ denote the diameter of $H$ (i.e., the longest distance
between any two nodes), and let
$K_x(H)$ denote the maximum number of connected components that
can be formed by removing at most $x$ nodes from $H$.
Note that if the diameter ${\rm diam}(H)$ is small, we can use
that value instead of $|H|$ in our neighborhood cover of $G$,
reducing the tree-width of the subgraphs $G_i$ to $O({\rm diam}(H))$.

\begin{lemma}\label{few-boundaries}
Let $H$ be a given pattern graph, and $N$ be a node of a
tree decomposition of graph $G$.  Then there are
at most $2^{K_{|L(N)|}(H) + |L(N)|\log(|H|+1)}$ different possible
partial isomorph boundaries $B:H\mapsto G'_N$ (as defined in the proof of
Lemma~\ref{dyn-prog-count}).
\end{lemma}

\begin{proof}
The map $B$ can be defined by specifying which (if any) vertex of $H$
maps to each vertex in $L(N)$
(using $\log(|H|+1)$ bits per vertex of $L(N)$ to specify this
information) and also specifying which of the remaining vertices in $H$
map to the vertex $x$ in $G'_N$ and which ones map to the vertex $y$.
However, it is not possible for the boundary to come from an actual
subgraph isomorphism unless each connected component of
$H\setminus B^{-1}(L(N))$ is mapped consistently either to $x$ or to $y$,
since any path from $x$ to $y$ must pass through a vertex of $L(N)$.
So, to finish specifying the boundary, we need only add this single bit
of information per component of $H\setminus B^{-1}(L(N))$,
and by definition there are at most $K_{|L(N)|}(H)$ such components.
\end{proof}

As a consequence, the analysis in Lemma~\ref{dyn-prog-count} can be
tightened to show that the dynamic program takes time
$2^{O(K_{O(diam(H))}(H)+{\rm diam}(H)\log |H|)}n$.

\begin{lemma}\label{few-components}
Suppose that planar graph $H$ is Hamiltonian,
or is 3-connected, or is connected and has bounded degree.
Then for any set $S$ of vertices of $H$,
$H-S$ has $O(|S|)$ connected components.
\end{lemma}

\begin{proof}
For Hamiltonian graphs and bounded degree graphs this is straightforward.
For 3-connected graphs, assume without loss of generality that
no edge can be added to $H$ connecting two vertices in $S$;
then each component of $H-S$ must occupy a distinct face in the planar
embedding of $S$ induced by the unique embedding of $H$.
\end{proof}

\begin{theorem}\label{Hamiltonian}
If a given pattern $H$ is Hamiltonian, 3-connected, or connected of bounded degree,
we can count the isomorphs of $H$ in a planar text graph $G$ with $n$
vertices in time $2^{O({\rm diam}(H)\log |H|)} n$.
\end{theorem}

\begin{proof}
The proof consists simply of plugging the improved analysis of
Lemma~\ref{few-boundaries} into the algorithm of Theorem~\ref{main}.
\end{proof}

For instance we can count the isomorphs of a wheel $W_k$ in a planar
text graph $G$ with $n$ vertices, in time $O(n k^{O(1)})$.  In fact in
this case it is not difficult to come up with an $O(nk^2)$ algorithm:

\begin{theorem}\label{wheel}
We can count the isomorphs of any wheel $W_k$ in a planar text graph $G$
with $n$ vertices in time $O(nk^2)$.
\end{theorem}

\begin{proof}
For each vertex $v$, we count the number of cycles of length $k$ in the
neighbors of $v$.  The sum of the sizes of all neighborhoods in $G$ is
$O(n)$.  Each neighborhood has treewidth at most~2 by
Lemma~\ref{diam-width}.  Any partial isomorphism of a $k$-cycle in a
node $N$ of this decomposition can only consist of a single path of
at most $k-1$ vertices, which starts and ends at some two of the at most
three vertices in $L(N)$ and may or may not involve the third vertex;
therefore we need only keep track of $O(k)$ different partial isomorph
boundaries at each node. A careful analysis of the steps in
the algorithm of Lemma~\ref{dyn-prog-count} then shows that the most
expensive step (finding compatible triples) can be performed in time
$O(k^2)$ per node, giving an overall running time of $O(k^2 n)$.
\end{proof}

\section{Variations and Applications}

We now describe briefly how to use our subgraph isomorphism algorithm to
solve certain other related graph problems. For instance, we can find the
girth (shortest cycle), or smallest nonfacial cycle (for some particular
embedding) simply by searching for isomorphs of small cycles:

\begin{theorem}
We can find the girth $g$ of any planar graph $G$,
or find a smallest nonfacial cycle $C$ for an embedded planar
graph, in time $2^{O(g\log g)}n$ or $2^{O(|C|\log|C|)}n$ respectively.
\end{theorem}

\begin{proof}
We test for each integer $i=3,4,5\ldots$ whether there is a cycle
(or nonfacial cycle) of length $i$.  To test if there is a nonfacial
cycle, we count the total number of cycles of length $i$ in the graph
and subtract the number of faces of length $i$.
The total time for this procedure is
$\sum_{i\le g} 2^{O(i\log i)}n=2^{O(g\log g)}n$.
For the nonfacial cycle problem, once the length of the cycle is known,
we can find a single such cycle by performing our subgraph isomorph
listing algorithm, stopping once $k+1$ cycles are generated, 
where $k=O(n)$ is the number of faces of the given length.
By radix sorting the list of cycles (in lexicographic order by their
sequences of vertex indices) we can then test in linear time which of
the generated cycles are nonfacial.
\end{proof}

\begin{theorem}\label{sep-cycle}
We can find a shortest separating cycle $C$ in a planar graph,
in time $2^{O(|C|\log|C|)}n$.
\end{theorem}

\begin{proof}
We describe how to test for the existence of a separating cycle of
length $i$; the shortest such cycle can then be found by a sequential
search similar to the computation of the girth.

We first modify the
construction of the graphs
$G_i$ in Lemma~\ref{nbhd-cover}, by including in $G_i$ not only all the vertices
in layers $S_i$ through $S_{i+w-1}$ but also a single supervertex
for each connected component of the graph induced by the vertices
in layers $S_j$, $j\ge i+w$, and a supervertex for the (single)
connected component of the vertices in layers $S_j$, $j<i$.
Then a cycle that uses only vertices in layers $S_i$ through $S_{i+w-1}$
(and does not use any of the supervertices) is separating in the
modified $G_i$ if and only if the corresponding cycle is separating in
$G$. Note that the added supervertices only add one level to the breadth
first search tree of $G_i$ and hence the tree-width is still $O(w)$.

Then we need merely modify the dynamic program of
Lemma~\ref{dyn-prog-count}, to use a definition of a partial isomorph
boundary that, in addition to the map $B:H\mapsto G'_N$, specifies which
of the remaining unmapped vertices of $G'_N$ are in each of the two
subsets of $G$ separated by the cycle, enforcing the
requirement that no vertex in one subset be adjacent to a
vertex in the other subset.  This modification multiplies the number of
boundaries by
$2^{O(w)}$, but this increase is swamped by the $2^{O(w\log w})$ terms
from our previous analysis.
\end{proof}

We next consider the application of our techniques to certain
graph clustering problems.

\begin{defn}[Keil and Brecht~\cite{KeiBre-CMCC-91}]
The {\em $h$-clustering problem} is that of approximating the maximum
clique by finding a set of $h$ vertices inducing as many edges as
possible.  The {\em connected $h$-clustering} problem adds the
restriction that the induced subgraph be connected.
\end{defn}

Keil and Brecht~\cite{KeiBre-CMCC-91} study these problems, and show that
even though cliques are easy to find in planar
graphs~\cite{PapYan-IPL-81}, the connected $h$-clustering problem is
NP-complete for planar graphs. See~\cite{KorPel-FOCS-93} for approximate
$h$-clustering algorithms in general graphs.

\begin{theorem}
For any $h$ we can solve the planar $h$-clustering and connected
$h$-clustering problems in time $2^{O(h\log h)} n$.
\end{theorem}

\begin{proof}
We simply to test subgraph
isomorphism for all possible planar graphs on $h$ vertices,
and return the subgraph isomorph with the most edges.
\end{proof}

We now describe two applications to connectivity that,
unlike the previous applications, are linear without an exponential
dependence on a separate parameter.

\begin{figure}[t]
$$\includegraphics[width=4.5in]{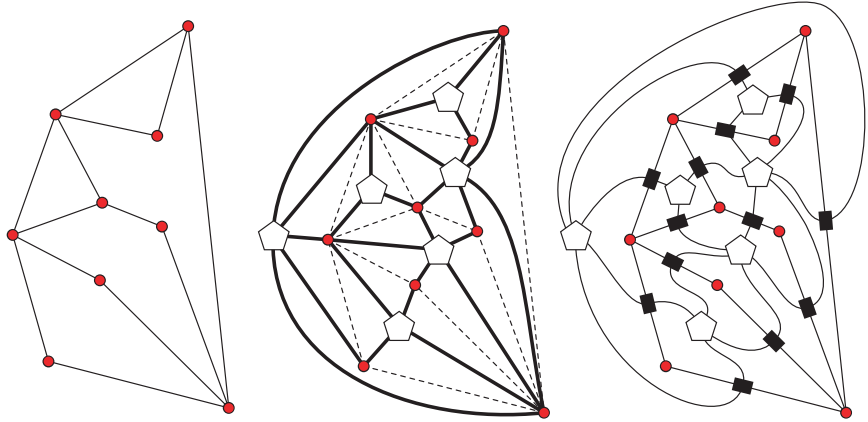}$$
\caption{An embedded planar graph $G$, the vertex connectivity substitute
graph $G'$ (with edges drawn as heavy curves), and the edge connectivity
substitute graph~$G''$.}
\label{fig:connsub}
\end{figure}

The {\em vertex connectivity} of $G$ is the minimum number of vertices
such that their deletion leaves a disconnected subgraph.  Since every
planar graph has a vertex of degree at most five, the vertex connectivity
is at most five.  We now use a method of Nishizeki (personal
communication) to transform the vertex connectivity problem into one of
finding short cycles, similar to those discussed at the start of this section.
We choose some plane embedding of $G$, and construct a new graph
$G'$ having $n+f$ vertices: the $n$ {\em original vertices} of $G$
and $f$ new {\em face-vertices} corresponding to the faces of $G$.
We place an edge in $G'$ between an original vertex and a face-vertex
whenever the corresponding vertex and face are incident in $G$.
Then $G'$ is a bipartite plane-embedded graph.
This construction is illustrated in the center
of Figure~\ref{fig:connsub}.

\begin{lemma}\label{vc-subst}
Any minimal set $S\subset G$ of vertices the deletion of which would
disconnect
$G$ corresponds to a cycle $C\subset G'$ of the same original vertices
and an equal number of face-vertices in $G'$,
such that $G'\setminus C$ has at least two connected components
each containing at least one original vertex.
Conversely if $C$ is any cycle such that $G'\setminus C$
has at least two connected components each containing at least one
original vertex, then the original vertices in $C$ form a cutset in $G$.
\end{lemma}

\begin{proof}
Let $A$ be a connected component of $G\setminus S$,
let $G_A$ be formed from $G$ by contracting $A$ into a supervertex,
and let $S'$ be the set of faces and vertices adjacent to
the contracted supervertex.
Then (since it is just the neighborhood of a vertex)
$S'$ has the structure of a cycle in $G'$,
and separates $A$ from $G\setminus\{A\cup S\}$.  If $S$ is minimal,
then it must consist of exactly the original vertices in $S'$.
The converse is immediate, since no edge in the embedding of $G$
can cross a face or vertex in $C$.
\end{proof}

\begin{theorem}
We can compute the vertex connectivity of a planar graph in
$O(n)$ time.
\end{theorem}

\begin{proof}
We can assume without loss of generality that $G$ is two-connected,
so the graph $G'$ described in Lemma~\ref{vc-subst} has no multiple
adjacencies.
We form $G'$ as above and find the shortest cycle in $G'$ that
separates two original vertices.
As with the shortest separating cycle problem (Theorem~\ref{sep-cycle})
this can be done by a slight modification to our dynamic programming
method that decorates the dynamic programming states with $O(1)$ bits of
additional information regarding the separated vertices.
Since any planar graph has a vertex of degree at most five by Euler's
formula, the shortest cycle in $G'$ must have length at most ten, so the
algorithm takes time $O(n)$.
\end{proof}

The {\em edge connectivity} of a graph is similarly defined
as the minimum number of edges the removal of which disconnects the graph.
For simple graphs, this can again be at most five
but for multigraphs it can be higher.

\begin{theorem}\label{edgecon}
We can compute the edge connectivity of a simple planar graph in
$O(n)$ time.
\end{theorem}

\begin{proof}
Assume without loss of generality that $G$ is two-edge-connected.
Embed $G$, and form a graph $G''$ by subdividing each edge of $G$
and connecting the resulting subdivision points to new vertices
in each adjacent face;
this construction is illustrated in the right
of Figure~\ref{fig:connsub}.
Then $G''$ is a planar graph with $n$ original vertices,
$e$ {\em edge-vertices} on each edge of $G$,
and $f$ {\em face-vertices} in each face of $G$,
so its total complexity is $O(n)$.
By the assumption of two-edge-connec\-ti\-vi\-ty, $G''$ is simple.
One can use an argument similar to the one in Lemma~\ref{vc-subst},
in which we delete a cutset from $G$, contract a connected component,
and examine the neighborhood in $G$ of the contracted supervertex,
to show that $G$ is $k$-edge-connected
iff there is no cycle of fewer than $2k$ edge- and face-vertices in $G''$,
which separates two original vertices. As before, the degree bound on a
planar graph imposes a limit of ten on the length of the shortest such
cycle, and as before this cycle can be found by a minor modification to
our dynamic programming algorithm.
\end{proof}

The same methods extend easily enough to multigraphs,
but now the edge connectivity can not be bounded a priori,
so we need to include the connectivity in our time bound.

\begin{theorem}
For any fixed $k$, we can test $k$-edge-connectivity of a planar
multigraph in time $2^{O(k\log k)}n$.
\end{theorem}

\begin{proof}
Without loss of generality, the multiplicity of any edge is at most $k$,
as higher multiplicities can not improve the overall connectivity.
After the edge subdivision step in the construction of $G''$,
the resulting graph is a simple planar graph with $O(kn)$ vertices, after
which we can proceed as in the remainder of Theorem~\ref{edgecon}.
\end{proof}

\section{Shortest Path Data Structure}

We next describe a technique for finding shortest paths in planar graphs.
Let a parameter $\ell$ be given (typically, a fixed constant).
We wish to test, for any two vertices $u$ and $v$,
whether there is a path from $u$ to $v$ of distance at most $\ell$,
and if so return the shortest such path.

Since we wish to use an amount of space independent of $\ell$,
we need a variant of Lemma~\ref{nbhd-cover} in which the total size of
the subgraphs is not so large.

\begin{lemma}\label{coarse-cover}
Let $G$ be a planar graph, and $w$ be a given integer parameter.  Then
we can find a collection of subgraphs $G_i$ and a partition of the
vertices of $G$ into subsets $S_i$ with the following properties:
\begin{itemize}
\item Every vertex of $G$ is included in at most two subgraphs $G_i$.

\item We can find a tree decomposition of each
subgraph $G_i$ with tree-width $O(w)$.

\item $G_i$ contains the $\ell$-neighborhood of every vertex in $S_i$.

\item The total time for performing the partition and computing the tree
decompositions is $O(w n)$.
\end{itemize}
\end{lemma}

\begin{proof}
As in Lemma~\ref{nbhd-cover}, we compute the distances of each vertex
from some arbitrary starting vertex $v_0$.
We then let $S_i$ consist of those vertices with distance at least
$2iw$ and at most $(2i+2)w-1$ from $v_0$,
and we let $G_i$ be the graph induced by the set of vertices
with distances at least $(2i-1)w$ and at most $(2i+3)w-1$ from $v_0$.
The proof that these graphs have treewidth $O(w)$ and that
each $G_i$ contains the $\ell$-neighborhood of $S_i$
is essentially the same as that of Lemma~\ref{nbhd-cover}.
\end{proof}

As before, by introducing dummy nodes, we can assume without
loss of generality that each node in the tree decomposition of $G_i$ has
at most three neighbors.
We warm up with a data structure for our shortest path queries that uses 
more space than necessary, but for which queries are very fast.

\begin{theorem}\label{fastq}
Given a planar graph $G$, and any value $\ell$,
we can in time and space $O(\ell n\log n)$ build a data structure
that can, given a query pair of vertices, either return the distance
between the pair or determine that the distance is greater than $\ell$,
in time $O(\ell)$ per query.
\end{theorem}

\begin{proof}
By performing the decomposition of Lemma~\ref{coarse-cover}
we can assume without loss of generality that we have a tree
decomposition $T$ for $G$ of width $O(\ell)$.
As with any tree, we can find a node $N$ the removal of which disconnects
$T$ into subtrees of size at most $|T|/2$.
Our primary data structure consists of the distances $d(x,z)$ from
each vertex $x\in G$ to each vertex $z\in L(N)$,
together with a recursively constructed data structure
in each subtree.

To answer a query pair $x,y$ where the two vertices
belong to different subtrees of $T$, we can simply try each
of the $O(D)$ values $d(x,z)+d(z,y)$ where $z$ ranges over
all the members of $L(N)$.
To answer a query where the two vertices belong to the same subtree,
we can use the recursively defined structure in that subtree.

It remains to show how we quickly determine which node $N$ is eventually
used to answer each query.  To do this, define the {\em level} of a node
to be the stage in the recursive subdivision process at which the node
was chosen, and define the {\em superior} of a node $N$ to be the node
chosen at the next earlier level in the subtree containing $N$.
The links from a node to its superior define a tree structure
$T'$ different from the original decomposition tree $T$. Further, define
the {\em home node} of a vertex
$v$ to be the node with the earliest level with $v\in L(N)$.
Note that, because of the requirements that the labels containing $v$
are contiguous, the home node is uniquely defined.
Then, the node to be used in answering a query pair $x,y$
is simply the least common ancestor in $T'$ of the home nodes of $x$
and~$y$.
\end{proof}

To return the actual shortest path, rather than simply the distance
between a pair of nodes, we can store a single-source shortest path tree
for each member of $L(N)$, and return the path in the tree for the
member of $L(N)$ giving the smallest distance.

We next show how to reduce the space to linear, at the expense of
increasing the query time.

\begin{theorem}
Given a planar graph $G$, and any value $\ell$,
we can build a data structure of size $O(\ell n)$
that can, given a query pair of vertices, either return the distance
between the pair or determine that the distance is greater than $\ell$,
in time $O(\ell^2\log n)$ per query.
\end{theorem}

\begin{proof}
As above, we can assume $G$ has a tree decomposition of width
$O(\ell)$, which we assume has the form of a rooted binary tree.
Define levels in this tree and home nodes of vertices as above,
except that we terminate the recursive subdivision process
when we reach subtrees with fewer than $\ell$ nodes
(which we call {\em small subtrees}).
If the node labels containing a vertex $v$ belong
only to nodes in a small subtree $S$, then $v$ does not
have a home node, instead we say that $S$ is $v$'s {\em home subtree}.

Define a pair of nodes in $T$ to be {\em related}
if there is no node between them with an earlier level than both.
Then, each node $N$ is related to $O(1)$ nodes at each level:
one node at an earlier level than $N$,
and at most one node in each later level
in each of the at most three subtrees formed by removing $N$ from $T$.
Therefore, there are $O(n/\ell)$ pairs of related nodes.

Our data structure then consists of the matrix of distances
from vertices in $L(N_1)$ to vertices in $L(N_2)$, for
each pair $N1,N2$ of related nodes.  The space for this data structure
is
$O(n\ell)$.  It can either be built as a subset of the data structure of
Theorem~\ref{fastq}, in time $O(n\ell\log n)$,
or bottom-up (using hierarchical clustering techniques of
Frederickson~\cite{Fre-SJC-97} to construct the level structure in $T$,
and then computing each distance matrix from two previously-computed
distance matrices in time $O(\ell^3)$) in total time $O(n\ell^2)$;
we omit the details.

To answer a query, we form chains of related pairs connecting the home
nodes (or small subtrees) of the query vertices to their common ancestor
in $T'$.  The levels of the nodes in these two chains becomes earlier at
each step towards the common ancestor, so the total number of pairs in
the chain is $O(\log n)$.  We then build a graph,
in which we include an edge between each pair of vertices in the labels
of a related pair of nodes, labeled with the distance stored in the
matrix for that pair.  We also include in that graph the edges of $G$
belonging to the small subtrees containing the query vertices, if they
belong to small subtrees.  The query can then be answered by finding a
shortest path in this graph, which has $O(\ell\log n + \ell^2)$ vertices
and $O(\ell^2\log n)$ edges.
\end{proof}

The following theorem on computing diameter improves the naive $O(n^2)$
bound for all pairs shortest paths when the diameter is small.
Note that diameter is not a subgraph isomorphism problem but it succumbs
to similar techniques.

\begin{theorem}\label{diameter}
We can compute the diameter $D$ of a planar graph $G$,
in time $O(2^{O(D\log D)} n)$.
\end{theorem}

\begin{proof}
We begin by performing a breadth first search from an arbitrary vertex.
This will produce a tree of height at most $D$, so by
Lemma~\ref{diam-width} we can find a tree
decomposition of width $O(D)$, which as usual we can assume has the
form of a rooted binary tree.  We first perform a bottom-up sweep of
this tree to compute for every node
$N$ the distances between every pair of vertices in $L(N)$,
in the graph associated with the subtree
rooted at $N$.  These $O(D^2)$ distances can be found by combining the
distance matrices of the two children of $N$, in time $O(D^3)$, so this
phase takes time $O(D^3 n)$.  We then sweep the decomposition top down,
computing for every node $N$ the distances between every pair of
vertices in $L(N)$, in the whole graph $G$.
The first pass correctly computed these distances at the root of the
tree, and at any other node $N$ the distances can be computed by
combining the distance matrices of the parent of $N$ (previously
computed in the top-down sweep) and its two children (computed in the
bottom-up sweep), again in time $O(D^3)$ per node.

We finally sweep through the tree decomposition bottom up,
keeping at each node $N$ a subset $S$ of the vertices seen so far in the
subtree rooted at $N$, together with the distances from each
member of $S$ to each member of $L(N)$. When we process a node $N$, we
perform the following steps:
\begin{enumerate}
\item Let the set $S$ for node $N$ consist of the union of the
corresponding sets $S_1$ and $S_2$ for its children $N_1$ and $N_2$,
together with
$L(N)$.
\item Compute the distances from each member of $S$ to
each member of $L(N)$, by combining the previously computed distances to
$L(N_1)$ or
$L(N_2)$ with the distances within $L(N_1)\cup L(N_2)$.
\item Compute the distance between each pair of nodes $x,y$
where  $x\in S_1\setminus L(N)$ and  $y\in S_2\setminus L(N)$,
by testing the distances through all $O(D)$ possible
intermediate nodes in $L(N)$.
\item Radix sort the members of $S$ according to the lexicographic
ordering of their $O(D)$-tuples of distances to $L(N)$,
and eliminate all but one member for each distinct tuple.
\end{enumerate}
The value returned as the diameter is then the maximum of the distances
from $S$ to $L(N)$ computed in the second step, and the distances
of pairs $x,y$ comuted in the third step

If any eliminated member $x$ of a tuple belongs to a diametral pair
$x,y$, where $y$ is not in the subgraph associated with $N$, then
the uneliminated member $x'$ with the same tuple would
have the same distance to $y$, and would form another diametral pair.
Therefore, the algorithm above will correctly find and report
a diametral pair.

The number of distinct $O(D)$-tuples of integers in the range from $0$
to $D$ is $2^{O(D\log D)}$, hence this gives a bound on the size of each
set $S$.  The time to compute distances between pairs $x,y$ is
$O(D)$ times the square of this quantity, which is still
$2^{O(D\log D)}$.
\end{proof}

\section{Other Graph Families}

Our results for planar graphs use the assumption of planarity in two
ways. First, in the bound relating tree-width to diameter
(Lemma~\ref{diam-width}), the proof is based on the existence of a
planar embedding of the graph, and in fact there is no similar bound in
general for nonplanar graphs; for instance the complete graph $K_n$ has
diameter one but tree-width $n-1$.
Second, in the cover of $G$ by low-treewidth subgraphs described
in Lemma~\ref{nbhd-cover}, we needed the fact that planar graphs are
closed under minors to show that the graph $G'_i$ is planar, allowing us
to apply Lemma~\ref{diam-width} to it.

This naturally raises the question, for which other minor-closed graph
families can we prove a bound relating diameter to tree-width, similar
to Lemma~\ref{diam-width}? Such a result would then let us apply our
subgraph isomorphism techniques unchanged to any such families. In the
conference version of this paper~\cite{Epp-SODA-95}, we announced an exact
characterization of these families, which are detailed in a separate
journal paper~\cite{Epp-Algo-?} and which we now summarize:

\begin{figure}
$$\includegraphics[width=4.5in]{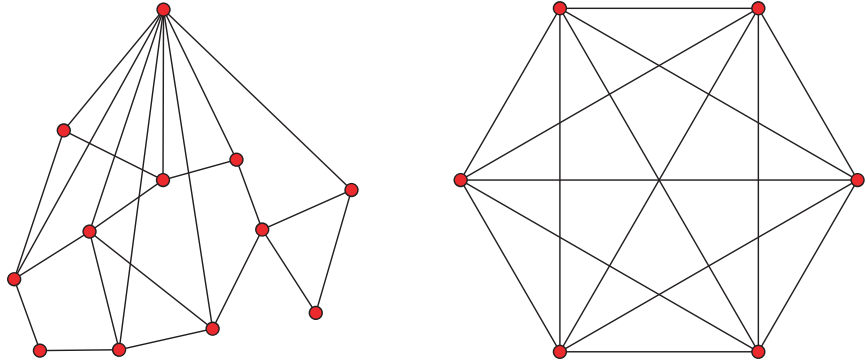}$$
\caption{The graph on the left is an apex graph; the topmost vertex is
one of the possible choices for its apex. The graph on the right
is not an apex graph.}
\label{fig:apex}
\end{figure}

\begin{defn}
Define a family $\cal F$ of graphs to have the {\em
dia\-me\-ter-tree\-width property} if
there is some function $f(D)$ such that every graph in $\cal F$ with
diameter at most $D$ has treewidth at most $f(D)$.
\end{defn}

\begin{defn}
An {\em apex graph} is a graph $G$ such that for some vertex~$v$
(the {\em apex}), $G\setminus\{v\}$ is planar (Figure~\ref{fig:apex}).
\end{defn}

\begin{theorem}[\cite{Epp-Algo-?}]
\label{dtmcgf}
Let $\cal F$ be a minor-closed family of graphs.
Then $\cal F$ has the diameter-treewidth property
iff $\cal F$ does not contain all apex graphs.
\end{theorem}

\begin{corollary}\label{apex-free-subiso}
Let $\cal F$ be a minor-closed family of graphs,
such that some apex graph is not in $\cal F$.
Let $H$ be a fixed graph in $\cal F$.
Then there is a linear time algorithm for testing
whether $H$ is a subgraph of a given graph $G\in\cal F$.
\end{corollary}

Note that the bound on tree-width from Theorem~\ref{dtmcgf}
is much higher than the linear bound in Lemma~\ref{diam-width},
so the dependence on $|H|$ of the time bound of the algorithm
implied by Corollary~\ref{apex-free-subiso}
is much greater than in our planar graph algorithms.
However, for certain important minor-closed graph families
(such as bounded genus graphs) we were able to prove a better
dependence of tree-width on diameter~\cite{Epp-Algo-?},
leading to less impractical algorithms.

\section{Conclusions and Open Problems}

We have shown how to solve planar subgraph isomorphism for any pattern
in time $O(n)$.  We have also solved certain related problems in similar
time bounds.  A number of generalizations of the
problem remain open:

\begin{itemize}
\item We have shown that we can solve planar subgraph isomorphism even for
disconnected patterns in time $O(n)$.  Can we list all
occurrences of a disconnected pattern in time
$O(n + k)$?

\item Bui and Peck~\cite{BuiPec-SJC-92} describe an algorithm for finding
the smallest set of edges partitioning a planar graph into two sets of
vertices with specified sizes; if the edge set has bounded size their
algorithm has cubic running time.  Can we use our methods to find such a
partition more quickly?

\item It seems possible that the recently discovered randomized coloring
technique of Alon et al.~\cite{AloYusZwi-JACM-95} can improve the
dependence on the size of the pattern from $2^{O(w\log w)}$ to
$2^{O(w)}$, but only for the decision problem of subgraph isomorphism. 
Can we achieve similar improvements for the counting and listing versions
of the subgraph isomorphism problem?
\end{itemize}

\section*{Acknowledgements}

The author wishes to thank Sandy Irani, George Lueker, and the anonymous
referees, for helpful comments on drafts of this paper.

\clearpage
\bibliographystyle{abuser}
\bibliography{subiso}
\end{document}